\documentclass{pasa}%

\title[Resolved observations of transition disks]{Resolved observations of transition disks}
\author[S. Casassus]{Simon Casassus$^1$\thanks{scasassus@u.uchile.cl}\\
\affil{$^1$Departamento de Astronom\'{\i}a, Universidad de Chile, Casilla 36-D, Santiago, Chile}}
\jid{PASA}
\doi{10.1017/pas.\the\year.xxx}
\jyear{\the\year}

\usepackage[authoryear]{natbib}
\bibpunct{(}{)}{;}{a}{}{,}
\usepackage{aas_macros}
\usepackage{hyperref} 
\hypersetup{colorlinks,citecolor=blue,linkcolor=blue,urlcolor=blue}

\begin{document}%
\begin{abstract}
Resolved observations are bringing new constraints on the origin of
radial gaps in protoplanetary disks.  The kinematics, sampled in
detail in one case-study, are indicative of non-Keplerian flows,
corresponding to warped structures and accretion which may both play a
role in the development of cavities.  Disk asymmetries seen in the
radio continuum are being interpreted in the context of dust
segregation via aerodynamic trapping. We summarise recent
observational progress, and also describe prospects for improvements
in the near term.
\end{abstract}
\maketitle%
\section{INTRODUCTION}
\label{sec:intro}


Circumstellar disks evolution is an aspect of stellar formation
\citep{ 1987ARAA..25...23S, 2001ARAA..39..549Z,
  Williams_Cieza_2011ARAA..49...67W, Dunham_2015ApJS..220...11D}. In
Class~II young-stellar-objects, the dissipation of the protostellar
envelope exposes a pre-main-sequence star surrounded by a gaseous
accretion disk. In this framework the spectral-energy-distributions
(SEDs) provide constraints on radial disk structure.  Central cavities
(and gaps) are inferred in so-called transition disks \citep[and
  pre-transitional disks,][TDs
  hereafter]{Espailat_etal_2007ApJ...670L.135E}, whose structure is
being refined with resolved observations.

This review provides a selection of structures seen in TDs, with an
emphasis on the radio domain. Insights on the origin of the cavities
based on TD demographics are summarised in the accompanying review by
\citet{Owen2015arXiv151206873O}. We caution that resolved observations
of TDs are affected by many biases, as they target the brightest
sources with the largest cavities, so that they cannot be extrapolated
to reflect the whole population \citep[which is for now studied by
  complete photometric surveys,
  e.g.][]{Cieza2012ApJ...750..157C,Cieza2015MNRAS.454.1909C}. Instead,
such resolved observations inform on the range of possible
astrophysical phenomena at work in disks.

The observational identification of abrupt disk inclination changes
(Section~\ref{sec:warps}) provides the information on disk orientation
required to interpret the dynamics of residual gas inside TD cavities
(Sec.~\ref{sec:gas}). The outer rings surrounding the central cavities
of most TDs resolved so far in the sub~mm continuum show non-axial
symmetry, with intensity contrasts from a few to $\sim$100
(Sec.~\ref{sec:traps}). Complex spiral patterns are seen in the outer
disks, which are also beginning to be resolved in molecular lines
(Sec.~\ref{sec:spirals}). The exciting observation of giant
protoplanet candidates opens an enormous research potential, since
their formation is expected to carve a radial gap, but their detection
in TDs is still far from systematic (Sec.~\ref{sec:planets}).  The
inferred warped accretion flows, the dramatic non-axial symmetry of
some outer disks, and the observed spiral patterns, motivate questions
on their origins, their connection to embedded protoplanets, and on
their role in disk evolution (Sec.~\ref{sec:conclusion}).

\section{WARPS} \label{sec:warps}

Warps, or variable inclination or orientation as a function of
stellocentric distance, have long been suspected to occur in
circumstellar disks. In this Section we compile evidence for warps in
young circumstellar disks in a broader context, with special attention
to TDs and the case of HD~142527.

\subsection{Debris disks}

Warps are seen in debris disks, or Class~III young stellar objects
that have dissipated the left-over primordial gas. This is the case of
the edge-on warp seen in $\beta$~Pic
\citep[][]{Golimowski2006AJ....131.3109G,
  Millar-Blanchaer2015ApJ...811...18M}. Another example is AU~Mic
\citep[][]{Wang2015ApJ...811L..19W, Boccaletti2015Natur.526..230B}, and perhaps also HD~110058
\citep[][]{Kasper2015ApJ...812L..33K}. The identification of warps in
debris disks requires an edge-on view, and the lack of gaseous
counterparts hampers reaching definitive conclusions on their
structure, and so distinguish continuous warps from superimposed disks.

\subsection{Indications of warps in gas-rich disks}  \label{sec:gasrichwarps}

Inner warps, or a tilted inclination at the disk center, could lead to
the photometric variations seen in some high-inclination (so close to
edge-on) Class II classical T-Tauri
stars\footnote{i.e. $\lesssim$1~$M_\odot$ stars undergoing accretion},
such as in RY~Lup \citep[][]{Manset2009AA...499..137M}, TWA~30,
\citep{Looper2010ApJ...714...45L} and AA~Tau
\citep{Bouvier2013AA...557A..77B}.  The light curve variability of
AA~Tau is well accounted for by a magnetically induced warp on scales
of a few stellar radii or $\lesssim$0.1~AU
\citep[][]{Esau2014MNRAS.443.1022E}, and part of the occulting
structure could extend beyond 1~AU
\citep[][]{Schneider2015A&A...584A..51S_AATau}. In the NGC~2264 open
cluster up to $\sim$40\% of young and gas-rich cTTs with thick inner
disks present AA~Tau-like variability
\citep[][]{Alencar2010A&A...519A..88A}, suggesting that such inner
warps represent a fairly common phase in the early evolution of
circumstellar disks \citep[][]{McGinnis2015A&A...577A..11M}.  Light
curve variability on weeks and months timescales, due to obscuring
material, has also been reported in the TDs T~Cha
\citep[][]{Schisano2009A&A...501.1013S} and GM~Aur
\citep[][]{Ingleby2015ApJ...805..149I}.

In TDs warps have also been hinted at in the form of
significant inclination changes in observations with different angular
resolutions. For example in GM~Aur, with a stellar mass
$\lesssim$1~$M_\odot$, \citet{Hughes2009ApJ...698..131H} propose a
central warp to explain the small change in disk position
angle\footnote{the intersection of the plane of the disk with that of
  the sky} of $11\pm2~\deg$ when comparing the major axis of the
sub~mm continuum sampled at 0.3~arcsec, with that of the CO(3-2)
kinematics\footnote{assuming Keplerian rotation} sampled at
2~arcsec. Likewise, \citet{Tang2012AA...547A..84T} explain that in
AB~Aur ($\sim$2~$M_\odot$), the inclination inferred from near-IR
interferometry \citep[$i\sim20~\deg$ within the central
  AU][]{Einser2004ApJ...613.1049E} is close to that sampled over 20~AU
scales, and lower than sampled in coarser beams \citep[$i\sim36~\deg$
  outside
  70~AU][]{Pietu2005AA...443..945P}. \citet{Hashimoto2011ApJ...729L..17H}
also conclude, from polarised-differential-imaging (PDI) in the
near-IR, that the inner regions of AB~Aur are warped, given the
varying inclinations in a double ring structure, dropping from $\sim
43~\deg$ to $\sim 27~\deg$ over 90 to 200~AU in
radius. \citet{Tang2012AA...547A..84T} relate the warped structure
inferred from their molecular line data with a residual infalling
envelope.

Yet another example of such inclination changes with angular scale is
seen in MWC~758 ($\sim$2M$_\odot$), where near-IR interferometry
\citep[with VLTI+AMBER,][]{Isella2008AA...483L..13I} yields an
inclination of 30-40$~\deg$ in the central AU, while SMA observations
in CO(3-2), with a 0.7~arcsec beam spanning over 100~AU, yield an
inclination of $\sim$21$~\deg$
\citep[][]{Isella2010ApJ...725.1735I}. Similarly, in HD~135344B (also
$\sim2 M_\odot$), the VLTI+MIDI inclination is $\sim$60$~\deg$
\citep[][]{Fedele2008AA...491..809F}, while CO(3-2) data suggests
11$~\deg$ \citep[][]{Dent2005MNRAS.359..663D}, and IR direct imaging
restricts $< 20~\deg$ \citep[HST+NICMOS][]{Grady2009ApJ...699.1822G}.

From an observational point of view, there is a degeneracy in the
line-of-sight kinematics due to variable inclination, as in a warp,
and the non-Keplerian flows expected from infalling gas, as identified
in TW~Hya and HD~142527 by \citet{Rosenfeld2012ApJ...757..129R,
  Rosenfeld2014ApJ...782...62R}.  In TW~Hya, excess CO emission at
high velocities, additional to that expected from axially symmetric
disk models, could be due to a warp, that may also account for the
faint azimuthal modulation in scattered light imaging reported by
\citep[][]{Roberge2005ApJ...622.1171R}. In HD~142527, the infalling
gas reported by \citet{Casassus2013Natur, Casassus2013rocks} could
instead be due to a warp. 

\subsection{The HD~142527 warp}  \label{sec:hd142warp}

\begin{figure}
\begin{center}
  \includegraphics[width=0.7\columnwidth,height=!]{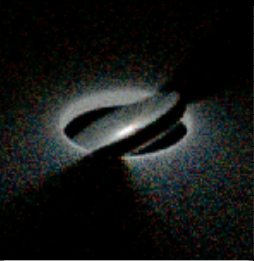}
\end{center}
  \caption{Figure adapted from \citet[][\copyright AAS, reproduced
    with permission]{Whitney2013ApJS..207...30W}, with a 3D radiative
    transfer prediction in $JHK$ for a warped disk configuration. The
    field is 4~AU on a side. }\label{fig:3Dwarp}
\end{figure}

The unambiguous identification of warps requires further evidence, in
addition to the possible inclination trends with angular scales (which
could be related to a variety of physical structures other than warps)
and the hints provided by the molecular line data. Radiative transfers
effects due to warps can provide such evidence. For instance in
HD~100546, \citet{Quillen2006ApJ...640.1078Q} recognised how a warped
structure could approximate the spiral pattern seen by HST
\citep[][]{Grady2001AJ....122.3396G}, thereby illustrating the
potential of radiative transfer effects to account for non-axially
symmetric structures. Indeed, \citet{Whitney2013ApJS..207...30W} apply
three-dimensional radiative transfer to warped disk architectures, as
shown in Fig.~\ref{fig:3Dwarp}, showing that shadows are expected in
the outer disks.

\begin{figure*}
\begin{center}
  \includegraphics[width=0.7\textwidth,height=!]{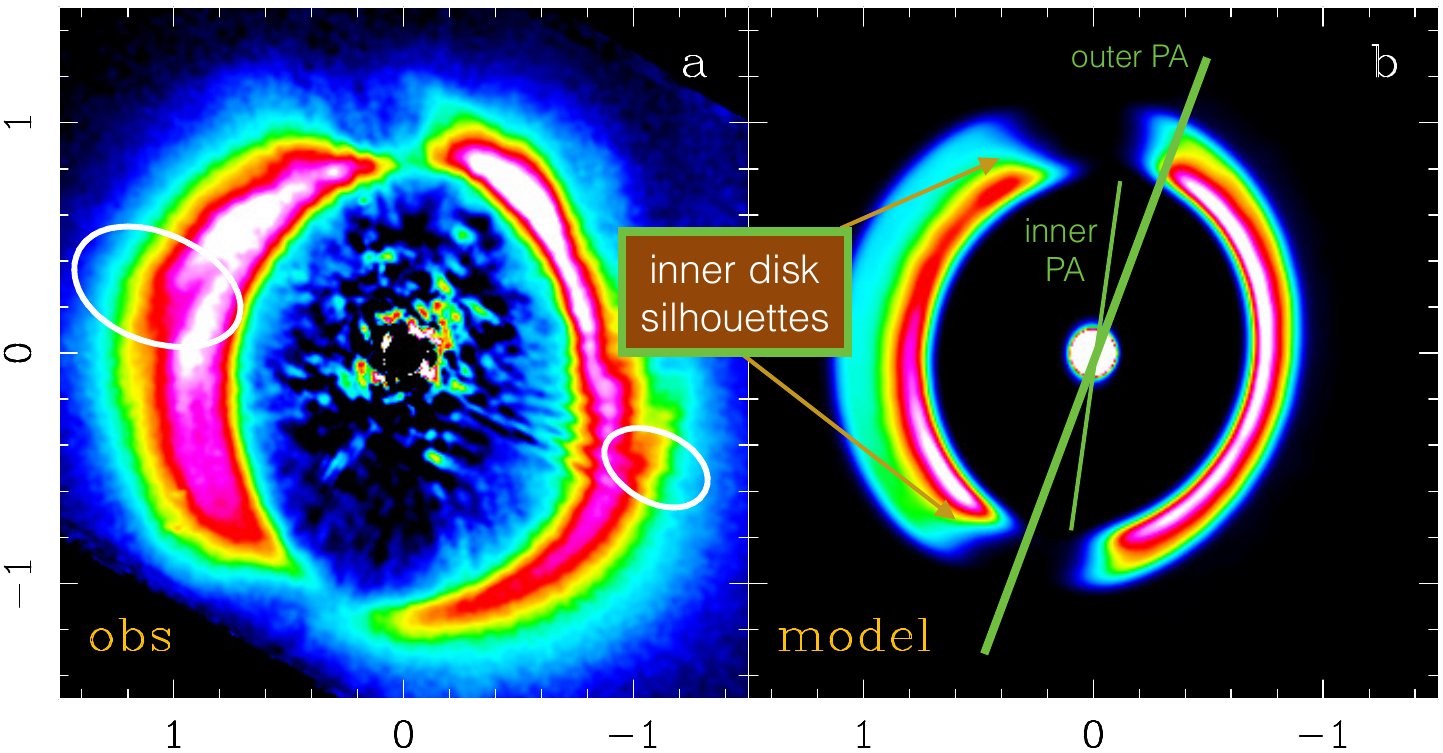}
\end{center}
  \caption{Comparison between the observed H-band polarised intensity
    image of HD~142527 \citep[{\bf
        a-} from ][]{Avenhaus2014ApJ...781...87A} and 3D radiative
    transfer predictions \citep[{\bf b-} from ][this is an updated
      version of their Fig.~2]{Marino2015ApJ...798L..44M}. The
    kinematics of C$^{18}$O(2-1) emission
    \citep[][]{Perez2015ApJ...798...85P} give the orientation of the
    outer disk; the white contours in {\bf a-} correspond to systemic
    velocities, so that the PA of the outer disk lies at
    $\sim160~\deg$ East of North, as indicated on {\bf b-}.  The inner
    disk shadows cast on the outer disk are best reproduced with a PA
    of $172~\deg$, the curvature of their outline (or silhouette) is
    reminiscent of the observations for the Eastern side. The
    similarities with the observations are particularly good given the
    idealisations of the model, which assumes a circular
    cavity.  \label{fig:pdiwarp} }
\end{figure*}

HD~142527 is an example of how such radiative transfer effects in
scattered light can unambiguously determine the disk orientation and
the existence of variable inclinations. In this case the relative
inclination change between the outer and inner disks reaches
$\sim70\pm5~\deg$, as shown by \citet{Marino2015ApJ...798L..44M}.
Despite such a dramatic inclination change, the intensity dips
originally identified by \citet[][]{Casassus2012ApJ...754L..31C}
eluded interpretation for 3 years, because the direction connecting
the dips is offset from the star, in contrast with that naively
expected for a simple tilt.  Yet, as illustrated in
Fig.~\ref{fig:pdiwarp}, the silhouette of the shadows cast by the
inner warp provide unequivocal evidence for this warp, which is
further supported by the gas kinematics discussed in
Sec.~\ref{sec:gas}.


\subsection{Near-term prospects: how common are warps?}

Although there are indications for the frequent occurrence of
inclination changes in gas-rich systems, there is so far only one
clear case of a warp, i.e. in HD~142527. Next-generation adaptive-optics
cameras, such as SPHERE and GPI, should soon provide substantial
improvements, particularly through polarised-differential-imaging
which appears ideally suited to trace illumination effects, and so
identify warps. First results from such new AO imaging are indeed
revealing intriguing intensity modulations, as in MWC~758
\citep[][]{Benisty2015AA...578L...6B}, HD~135344B
\citep[][]{Wahhaj2015AA...581A..24W} and in HD~100453
\citep[][]{Wagner2015ApJ...813L...2W}.  Followup molecular-line
observations with ALMA may provide the necessary clues from the gas
kinematics.

\section{GAS IN CAVITIES} \label{sec:gas}

Models predict that young giant protoplanets carve a deep gap in the
dust component of protoplanetary disks, and a shallower gap in the gas
and very small grains (VSGs) components
\citep[e.g.][]{PaardekooperMellema2006, Fouchet2010}.  The clearing of
the protoplanetary gap is thought of as an important mechanism
underlying the class of TDs, and residual gas in cavities
has thus been intensely searched for.

This Section describes observations of intra-cavity gas that select
sufficiently large cavities to be resolved with current gas tracers,
so with radii greater than a dozen AUs at the very least. Smaller
cavities and inner disks have been detected by broad-band
long-baseline optical interferometry \citep[e.g. the gaps in HD~100546
  or HD~139614][]{Tatulli2011A&A...531A...1T,
  Matter2016A&A...586A..11M}. Such observations do not yet sample the
intra-cavity gas \cite[except for the central few stellar radii in
  Br$\gamma$, e.g.][]{Mendigutia2015MNRAS.453.2126M}. The continuum
interferometric observations have allowed to isolate the crystalline
spectrum of the inner disks \citep[as in
  HD~142527][]{vanBoekel2004Natur.432..479V}, and to study the
occurence of gaps in HAeBes \citep[][]{Menu2015A&A...581A.107M}.

\subsection{Ro-vibrational gas and very small grains}

Long-slit spectroscopy has provided indirect evidence for gas inside
cavities, under the assumption of azimuthal symmetry and Keplerian
rotation \citep{Carr2001ApJ...551..454C, Najita2003ApJ...589..931N,
  Acke2006AA...449..267A, vanderPlas2008AA...485..487V,
  Salyk2009ApJ...699..330S}.  Spectro-astrometry has also been used to
infer a residual gas mass from ro-vibrational CO emission
\citep{Pontopiddan2008,vanderPlas2009AA...500.1137V,
  Pontoppidan2011ApJ...733...84P}. A comprehensive analysis is
presented in \citet{vdP2015A&A...574A..75V} and
\citet{Banzatti2015ApJ...809..167B}.  The general picture is that dust
cavities do indeed contain some amount of residual gas, as expected in
the context of dynamical clearing.

The smallest grains are thought to be well coupled with the gas, so
that they can be used as proxies. Thermal IR observations suggest the
existence of smaller gaps in VSGs than in silicates, and have also led
to the finding of gaps otherwise undetectable through the SED alone
\citep[][]{Maaskant2013A&A...555A..64M,
  Honda2015ApJ...804..143H}. Detailed studies of the gap structure in
ro-vibrational lines along with infrared data have been performed in
HD~135344B \citep[][]{Garufi2013AA...560A.105G, Carmona2014AA...567A..51C}.

\subsection{Resolved  radio observations of intra-cavity gas} \label{sec:resolvedcavity}

\begin{figure}
\begin{center}
  \includegraphics[width=0.8\columnwidth]{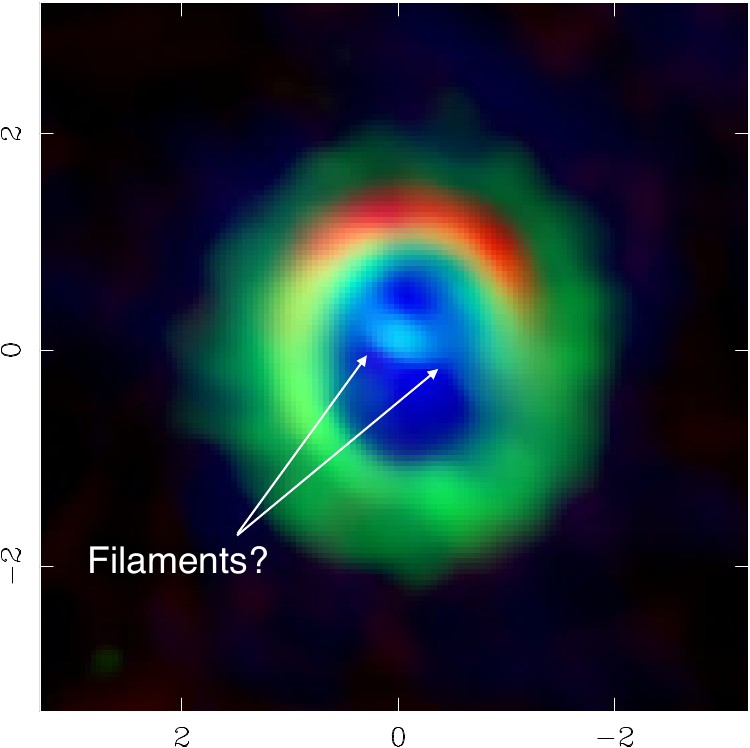}
  \end{center}
  \caption{Summary of Cycle~0 band~7 observations, from MEM maps, with
    continuum in red, HCO$^+$(4-3) in green, and CO(3-2) in blue
    \citep[adapted from][]{Casassus2013Natur}.  $x-$ and $y-$ show
    offset from the star, in arcsec. Velocities have been restricted
    to highlight the fainter structures seen in HCO$^+$, which are
    otherwise dwarfed by the fast HCO$^+$ central emission.
  \label{fig:ALMARGB}}
\end{figure}

While the ro-vibrational detections pointed at residual intra-cavity
gas, surprises were nonetheless brought by the first resolved images
\citep[Fig.~\ref{fig:ALMARGB},][]{Casassus2013Natur}, made possible
thanks to the advent of ALMA. The finest angular resolutions from ALMA
Cycle~0 were just sufficient to resolve the largest TD cavity,
i.e. that of the HD~142527 disk. A few $M_\mathrm{jup}$ worth of gas
was indeed seen inside this cavity from CO isotopologues
\citep[][]{Perez2015ApJ...798...85P}. However, the gas kinematics were
puzzling, and seemed consistent with radial infall as illustrated by
the fast and centrally peaked HCO$^+$(4-3).  Slower intra-cavity
HCO$^+$ signal seemed to connect with the outer disk along filamentary
structures, reminiscent of protoplanetary accretion streams. But the
slow intra-cavity HCO$^+$ was faint in these Cycle~0 data, and the
cavity radius was sampled only with 3~beams. New observations in
HCO$^+$(4-3) are required to ascertain its structure.

\begin{figure}
  \includegraphics[width=\columnwidth,height=!]{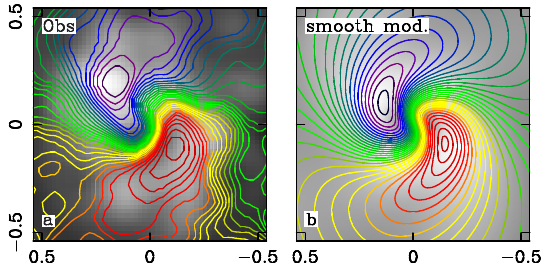}
  \caption{Figure adapted from \citet[][part of their Fig.~26, \copyright AAS, reproduced
      with permission]{Casassus2015ApJ...811...92C} , showing the
    observed CO(6-5) kinematics in the central regions of HD~142527,
    and comparing with model predictions for the accretion through the
    warp represented in Fig.~\ref{fig:cartoon} (convolved at the
    resolution of the centroid map). The origin of coordinates is set
    to the stellar position.  } \label{fig:co65warp}
\end{figure}

\begin{figure}
\begin{center}
\includegraphics[width=\columnwidth,height=!]{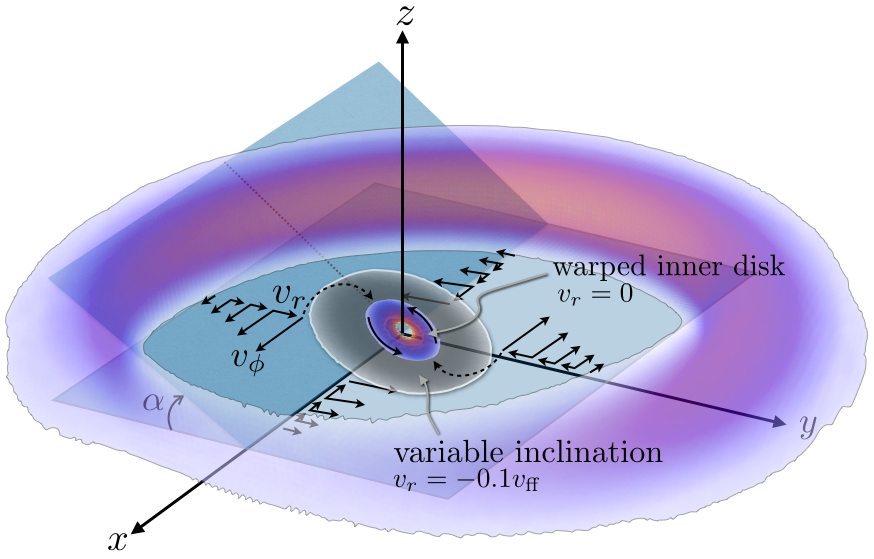}
\end{center}
\caption{Sketch of the accretion kinematics through the warp in
  HD~142527 \citep[from][]{Casassus2015ApJ...811...92C}.  The crescent in hues of
  red and purple represents the distribution of mm-sized grains.
  Inside the warp, as it connects the two disk inclination material
  flows along the dashed and curved arrows. \label{fig:cartoon}}
\end{figure}

A more accurate view of the intra-cavity kinematics in HD~142527 is
provided by the CO(6-5) line, with a higher frequency and a finer
Cycle~0 beam.  CO(6-5) is also free of the interstellar absorption
that affects CO(3-2) \citep[][]{Casassus2013AA...553A..64C, Casassus2013rocks}.  Given
the warped structure of HD~142527, the line-of-sight CO(6-5) can be
understood as infall, at the observed stellar accretion rate
\citep{2006AA...459..837G}, but along and through the warp. As
illustrated in Figs.~\ref{fig:co65warp} and Fig.~\ref{fig:cartoon},
the observations are fairly well accounted for by a model with gas
inside the continuous warp. At the radius where the disk plane crosses
the sky, the kinematics are Doppler-flipped so that blue turns to red,
and convolution with the angular resolution results in the
``S''-shaped centroids. Interestingly, a fast and narrow warp improves
the model, such that the two disk orientations are connected within
$\sim$3~AU, at a radius of 20~AU, with material flowing orthogonal to
the local disk plane at a velocity comparable to Keplerian. These
observations are based on non-parametric image synthesis that pushes
the limits of the Cycle~0 beam, so that further observations are
required to ascertain the details of the gaseous flows inside the
warp.

A likely origin for the observed accretion dynamics in HD~142527
probably involves the low-mass companion, at $\sim$12~AU and with a
mass ratio $\lesssim 1/10$ \citep[][]{Biller2012, Close2014,
  Rodigas_2014ApJ...791L..37R}. The proper-motion of the companion
\citep[][]{Lacour2015arXiv151109390L} seems to indicate an orbit in
clockwise rotation, as for the whole disk. If HD~142527B is contained
in the tilted inner disk, or if it is close to it, the properties of
the warp caused by the HD142527A+B system resemble disk tearing in
strongly warped circumbinary disks
\citep[][]{Nixon_2013MNRAS.434.1946N,
  2015MNRAS.448.1526N,2015MNRAS.449.1251D}.  In the context of
circumstellar disks, \citet{Facchini2013MNRAS.433.2142F,
  Facchini2014MNRAS.442.3700F} show that in the thick-disk regime,
where $\alpha < H/R$, a inner warp greater than $\sim$40~deg will
break the disk into distinct planes. Note, however, that the low mass
companion at $\sim$12~AU is unlikely to have originated the 140~AU
cavity, unless it is undergoing oscillations in inclination and
eccentricity \citep[i.e. Kozai cycles,
  see][]{Martin2014ApJ...792L..33M}.  Further progress in the
interpretation of the HD~142527 warp requires better data and tailored
hydrodynamical models.

\subsection{Residual gas in cavities}

A summary of several other detections with ALMA of residual gas in
TD cavities is given by
\citet{vandishoeck_2015arXiv150501947V}.  Bright CO(6-5) emission was
found ascribed to the cavities of HD~135344B and SR21
\citep[][]{Perez_L_2014ApJ...783L..13P}, as well as in LkCa~15,
RXJ~1615-3255, and SR~24S
\citep[][]{vandermarel_2015AA...579A.106V}. The size of the ring in
      [PZ99]J160421.7-213028 is smaller in CO(2-1) than in the dust
      continuum \citep[][]{Zhang2014ApJ...791...42Z}, much as in
      SZ~91\citep[][]{Canovas2015ApJ...805...21C}. The depth of the
      cavity in mm-sized grains follows from the sub~mm
      continuum. Results so far are limited by dynamic range, with
      depths in the mm-sized dust deeper than $\sim$1/100 relative to
      the outer disk.

The recent state of the art 345~GHz continuum and CO(3-2) isotopologue
data analysed by \citet{vdMarel_2016A&A...585A..58V} in terms of a
thermochemical model, suggests that the gas cavities are up to 3 times
smaller than the dust rings in DoAr~44, SR~21, HD~135344B and IRS~48
\citep[see also][]{Bruderer_2014AA...562A..26B}. The drop in gas
surface density can be up to $\delta \sim 10^{-2}$ \citep[the exact
  values vary in each object and can be calculated from the parameters
  given in Table~3 of][]{vdMarel_2016A&A...585A..58V}.

The main result that transpires from the above studies, including
HD~142527 \citep[][]{Perez2015ApJ...798...85P}, is that the depth of
the cavity is shallower in the gas than in the dust, by at least a
factor of 10. However, while \citet{vdMarel_2016A&A...585A..58V}
investigate gradual cavity edges, angular resolution limits prevent
firm conclusions on the distribution of the gas.  Most studies assume
axially symmetric and sharp cavity edges, parametrised by a
step-function drop, with a depth $\delta \sim 1/ 10 - 1/100 $ relative
to the outer disk in the gaseous component. Due to the coarse linear
resolutions available, the accretion kinematics remain largely
unsampled. For instance, in HD~142527 sufficient linear resolution is
crucial to trace the non-Keplerian accretion flows, which are
otherwise much less conspicuous when sampled with a beamwidth
comparable to the cavity radius \citep[][]{Perez2015ApJ...798...85P}.

\subsection{Near-term prospects: structure and origin of the cavities}

The finest angular resolutions available from the latest ALMA results
\citep[][]{vdMarel_2016A&A...585A..58V} do no yet provide enough
linear resolution to sample the cavity edges.  However, at the time of
writing several ALMA projects are currently underway in gas-rich
cavities, and the first results from the next-generation AO cameras
are soon due. With much finer angular resolutions, it will be possible
to understand better the distribution of the gas in TDs
and its kinematics, and so reconcile mass transport across TD cavities
with the stellar accretion rates (which has so far only been possible
in HD~142527).

A key question is on the shape of the cavity edge, which is currently
modelled as a step function. Yet the edge of the cavity gives
important clue on the clearing mechanism: for instance, if planet
formation, the sharpness of the edge is a function of the mass of the
outermost body \citep[e.g.][]{Crida2006Icar..181..587C,
  Mulders2013A&A...557A..68M}.

Intriguingly very few companions have been found in TDs
(and only one candidate near the edge of a cavity, i.e.  in the
HD~169142 disk, see Sec.~\ref{sec:planets}).  A spectacular example is
HD~142527, with a record-sized cavity of 140~AU and no intra-cavity
protoplanet detected as yet. The binary separation in HD~142527 is
$\sim$10~AU (see Sec.~\ref{sec:resolvedcavity}) and could only account
for a $\sim$30~AU gas-free cavity
\citep[][]{Artymowicz_Lubow_1994ApJ...421..651A}. A interesting
comparison object is the GG~Tau circumbinary ring, which by contrast
to HD~142527 may be explained in terms of dynamical interactions with
the binary provided the edge of the cavity is `soft' in the gas
\citep[][]{Andrews2014ApJ...787..148A, Dutrey2014Natur.514..600D}.

Despite the lack of detections in TD cavities, the data in HD~142527
suggest that low-mass companions can drive dramatic warps. Thus a
possible solution to the puzzle of large cavities could perhaps be
found in viewing them as circumbinary disks with low mass companions
undergoing Kozai oscilations. so that when the system is coplanar the
binary is very eccentric \citep[as suggested
  in][pending tests with hydrodynamic simulations]{Casassus2015ApJ...811...92C}. Thus in HD~142527 we could be
witnessing a high-inclination phase at relatively close binary
separation. Definitive conclusions on the structure of this warp
require finer angular resolutions and deeper sensitivities, along with
new observations in the HCO$^+$ lines. Here we are presented with
another puzzle to reconcile models of disk tearing with the
indications that even though the warp appears to be abrupt, gas
continuously flows through it and connects the two disk inclinations.

Is there a binary lurking in all of the large TD cavities, as in
HD~142527? Low mass companions into the brown-dwarf mass regime are
difficult to detect at very close separations. The new AO cameras
along with ALMA gas kinematics should soon cast light on these
questions, starting with the recent detection of fairly high-mass
bodies in LkCa~15 \citep[][see
  Sec.~\ref{sec:planets}]{Sallum2015Natur.527..342S}.


\section{LOPSIDED DISKS} \label{sec:traps}

Early resolved observations of TDs already suggested that the outer
rings deviate from axial symmetry \citep[in general from SMA
  data,][]{Ohashi2008Ap&SS.313..101O, Brown2009ApJ...704..496B,
  Andrews2011ApJ...732...42A}. ALMA has brought confirmation and
revealed the dramatic contrast\footnote{the ratio of the intensity
  extrema along a ring} of the most extreme asymmetries seen in the
dust continuum. To what extent do these asymmetries trace matched
asymmetries in the gas? What is the efficiency of dust segregation
from the gas?

\subsection{The most dramatic asymmetries}

\subsubsection{HD~142527}

The sub~mm continuum from HD~142527 is approximately shaped into a
crescent \citep{Ohashi2008Ap&SS.313..101O, Casassus2013Natur,
  2013PASJ...65L..14F} and modulated by a temperature decrement under
the shadow of the inner warp (Sec.~\ref{sec:hd142warp}). It is still
not entirely clear to what extend is the sub~mm continuum reflecting a
similarly pronounced asymmetry in the gas. Existing observations rely
on optically thick tracers of the total gas mass (including small dust
grains), so cannot conclude on the underlying gas distribution. The
dense-gas tracers discussed by \citet[][HCN(4-3) and
  CS(7-6)]{vanderplas2014ApJ...792L..25V} are affected by chemistry 
and seem to anti-correlate with the continuum.

\begin{figure*}
\begin{center}
\includegraphics[width=0.8\textwidth,height=!]{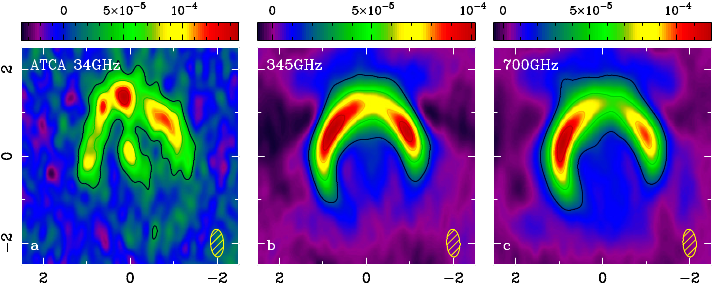}\\
\includegraphics[width=\textwidth,height=!]{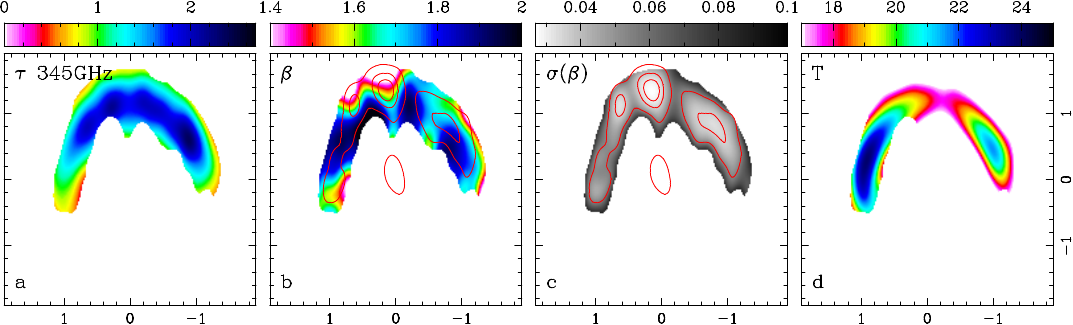}
\end{center}
\caption{ Figure adapted from \citet[][]{Casassus2015ApJ...812..126C} . {\bf Upper row:}
  multi-frequency data of HD~142527 brought to a common
  $uv$-coverage. The ALMA data have been filtered for the ATCA
  response with Monte-Carlo simulations of ATCA observations on
  deconvolved models of the ALMA data.  \textcolor{blue}{\bf a)}:
  restored ATCA image at 34~GHz.  \textcolor{blue}{\bf b)}: average of
  Monte-Carlo (MC) simulations of ATCA observations on the ALMA band~7
  data. \textcolor{blue}{\bf c)}: MC simulations of ATCA observations
  on the ALMA band~9 data.  {\bf Lower row:} Grey-body diagnostics
  inferred from the multi-frequency data.  \textcolor{blue}{{\bf a}}:
  optical depth map at the reference frequency of
  345~GHz. \textcolor{blue}{{\bf b}}: line of sight emissivity index
  map $\beta$, with ATCA specific intensity contours in red.
  \textcolor{blue}{{\bf c}}: root-mean-square uncertainties on the
  emissivity index map.  {\bf d}: line of sight temperature,
  $T_s$. \label{fig:Ttaubet} }
\end{figure*}

Recently the Cycle~0 data in the optically thin CO isotopologues were
further analysed by \citet{Muto2015PASJ...67..122M}, who estimate that
the contrast in the gas-phase CO column is about a factor of 3. While
this is all the existing information available on the gas contrast, at
the time of writing, its extrapolation to the total gas mass is
hampered by uncertainties from grain surface chemistry: The azimuthal
structure in these transitions is not a smooth crescent; their
emission peaks differ from the continuum. Perhaps the gas-phase CO
abundance is significantly modulated in azimuth by chemistry, or a
fraction of CO could be depleted on dust grains, especially at the
location of the continuum peak (at about 11h), where continuum
grey-body temperatures are $\sim$22~K
\citep[Fig.~\ref{fig:Ttaubet},][]{Casassus2015ApJ...812..126C}, so
below freeze-out \citep[][]{Jorgensen2015AA...579A..23J}.

The hydrodynamic simulations tend to predict contrast ratios of order
$\sim$3. Large-scale crescents in the gas surface density arise
naturally in models of cavity clearing by giant planets
\citep[e.g.][]{Zhu_Stone_2014ApJ...795...53Z}.  Such crescents are
reproduced in hydrodynamical simulations by large-scale anti-cyclonic
vortices, which result from Rossby wave instabilities triggered at
sharp radial gradients in physical conditions. For instance, they have
also been modeled in the context of viscosity gradients
\citep{2012MNRAS.419.1701R}.  An apparently different model for such
crescents, leading to more pronounced asymmetries in the gas, was
proposed by \citet[][]{Mittal2015ApJ...798L..25M} based on a global
disk mode in response to a stellar offset.  However, as argued by
\citet{ZB_2015arXiv151103497Z} if such offsets result from the shift
of the system centre of mass in sufficiently massive disk, the
consistent incorporation of disk self-gravity dampens the contrast of
an otherwise standard large-scale anticyclonic vortex. Thus in general
the gas surface density contrasts are predicted to reach moderate
values, typically $\sim$3 and perhaps up to 10.

The azimuthal intensity contrast in HD~142527 reaches about 30, yet
there is no hint so far of a similar contrast ratio either in the
small grains ($\ll$ 0.1~mm), nor in molecular lines, which led
\citet{Casassus2013Natur} to suggest that the mm-sized dust grains
that originate the ALMA continuum were segregated from the gas. In
other words the dust-to-gas mass ratio could vary with azimuth. A
promising mechanism to explain such segregation is the pile-up of
larger grains in local pressure maxima
\citep[e.g.][]{W1977MNRAS.180...57W, BS1995AA...295L...1B,
  Birnstiel2013AA...550L...8B, LyraLin2013ApJ...775...17L}, as
proposed to account for the even more dramatic sub~mm continuum
contrast seen in IRS~48 \citep[see
  below,][]{vanderMarel2013Sci...340.1199V}.

An important prediction of dust trapping is that progressively larger
grains\footnote{up to Stokes numbers of $\sim$1} should be more
sharply confined. Multi-frequency radio observations may test this
prediction, since grains of progressively larger size dominate the
continuum emission at correspondingly longer wavelengths. Indeed,
34~GHz ATCA observations of HD~142527 revealed a compact clump of
cm-wavelength-emitting grains buried into the ALMA crescent
\citep[Fig.~\ref{fig:Ttaubet},][]{Casassus2015ApJ...812..126C}. Populations
of larger grains correspond to shallower optical depth spectral index
$\beta$ \citep[e.g.][]{Testi2014prpl.conf..339T}, in a parametrisation
of the optical depth spectrum such that $\tau = \tau_\circ \times (\nu/\nu_\circ)^{\beta_s}$ ($\beta$ is also referred to as emissivity
index). Thus three frequency points can be used to solve for the
radiation temperature, optical depth, and $\beta$ index at each line
of sight.  Frequencies above $\sim$345~GHz turned out to be optically
thick \citep[Fig.~\ref{fig:Ttaubet},][]{Casassus2015ApJ...812..126C},
so that the spectral index trends at ALMA frequencies are due to
optical depth effects rather than the underlying dust population. The
ATCA clump translates into a local minimum in the $\beta$ index,
indicative of larger grains.

Interestingly the optical depth at 345~GHz (Fig.~\ref{fig:Ttaubet}a,
bottom), shows an extension towards 2h. Yet the minimum in opacity
spectral index $\beta$ lies at 11h, which should be the core of the
pressure maximum in the dust trap scenario. Since rotation in
HD~142527 goes clockwise, a possible interpretation lies in the
phenomenon predicted by \citet{BZ_2015arXiv151103498B}, that larger
grains should concentrate significantly ahead of the dust trap. Thus
at 2h the line of sight would intercept a disk-averaged grain
population, modulated by the gaseous crescent, but the addition of
larger grains (perhaps up to 1~cm sizes in this disk), might raise the
total optical depth even away from the pressure maximum. However, such
a shift of the larger grains is possible only with Stokes numbers $S
\gtrsim 1$ \citep{BZ_2015arXiv151103498B}, which for $1~$cm grains
would require a disk mass $\sim$100 times lower than inferred from the
SED modeling \citep[][]{Casassus2015ApJ...812..126C}, and/or a
correspondingly low gas-to-dust mass ratio.

Thus the bulk of the data so far indicates that the spatial
segregation of dust sizes in azimuth is indeed at work in the lopsided
outer ring of HD~142527. However, while models have been proposed that
approximate the observed segregation \citep[through parametric
  models,][]{Casassus2015ApJ...812..126C}, they are fine-tuned in key
parameters such as the $\alpha$ turbulence prescription and the
threshold grain size for trapping.

\subsubsection{IRS~48}

The record-holder lopsided TD is IRS~48, where
\citet[][]{vanderMarel2013Sci...340.1199V} found that the contrast in
ALMA band~9 is greater than 100, so three times as pronounced as in
HD~142527. They reproduce such dramatic contrast with their dust
trapping prescriptions. 

Observations of IRS~48 are hampered by intervening cloud emission
from $\rho$~Oph, with $A_\mathrm{V} \sim 10$.  Yet by using the rarer
CO isotopologues at higher velocities,
\citet{Bruderer_2014AA...562A..26B} estimated that the total gas mass of the
disk is only $\sim 10^{-4}~$M$_\odot$. 
 
\begin{figure}
\begin{center}
\includegraphics[width=0.8\columnwidth,height=!]{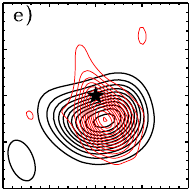}
\end{center}
\caption{Figure adapted from \citet[][their Fig.~ 1, \copyright AAS,
    reproduced with permission]{vandermarel_2015ApJ...810L...7V}.  VLA
  observations of IRS~48 at optically thin frequencies confirm the
  extreme lopsidedness of this TD.  VLA observations at 34~GHz are
  shown in red contours - the tail towards the NE is likely due to
  stellar emission.  ALMA band~9 observations at 680~GHz are shown in
  black contours, after filtering for the VLA response.  The field is
  2~arcsec on a side.  While at 34~GHz the crescent appears to be
  somewhat more compact, the role of optical depth effects in widening
  the 680~GHz signal remain to be quantified.  \label{fig:irs48}}
\end{figure}

The dramatic contrast observed in IRS~48 would seem very unlikely to
correspond to an equally lopsided gas distribution, and is thus
probably due to segregation of grain sizes. Indeed,
\citet{vandermarel_2015ApJ...810L...7V} report on the tentative
detection of the corresponding spectral trends in a comparison between
VLA and ALMA observations.  The confirmation of lopsidedness at
optically thin VLA frequencies (see Fig.~\ref{fig:irs48}) dissipates
worries that the asymmetry observed at high frequencies, which are
likely optically thick, could be related to optical depth effects in
this highly inclined disk.

\subsubsection{MWC~758}

\begin{figure}
\begin{center}
\includegraphics[width=\columnwidth,height=!]{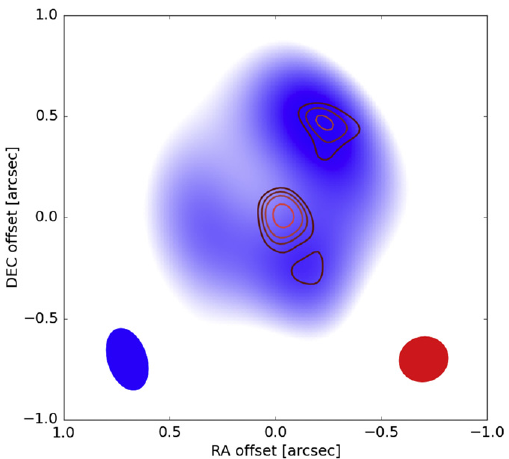}
\end{center}
\caption{Figure adapted from \citet[][their Fig.~ 3]{Marino2015ApJ...813...76M}.  VLA
  observations of MWC~758 compared to a deconvolved model of the ALMA
  band~7 visibilities. The VLA signal is shown in red contours - the
  red ellipse corresponds to the synthesised beam. The blue scale
  corresponds to an `MEM' non-parametric model of the 337~GHz ALMA
  data - the blue ellipse corresponds to a elliptical Gaussian fit to
  the point-spread-function of the `MEM'
  algorithm.  \label{fig:mwc758data}}
\end{figure}

Can all continuum asymmetries in TDs be interpreted as
dust traps? MWC~758 presents an interesting caveat. The VLA~34~GHz
image presented by \citet[][reproduced in
  Fig.~\ref{fig:mwc758data}]{Marino2015ApJ...813...76M}, when compared
with the more extended signal seen by ALMA, seems fairly consistent
with the dust trap scenario, at least as well as in HD~142527 or
IRS~48 given the available constraints. But in fact two pressure
maxima (or anticyclonic vortices) would seem to be required to account
for the double-peaked ALMA signal in MWC~758. Besides estimates of the
sub~mm optical depth are missing - which requires finer angular
resolution sub~mm data, as well as an additional frequency point
intermediate with the VLA.

\begin{figure*}
\begin{center}
\includegraphics[width=\textwidth,height=!]{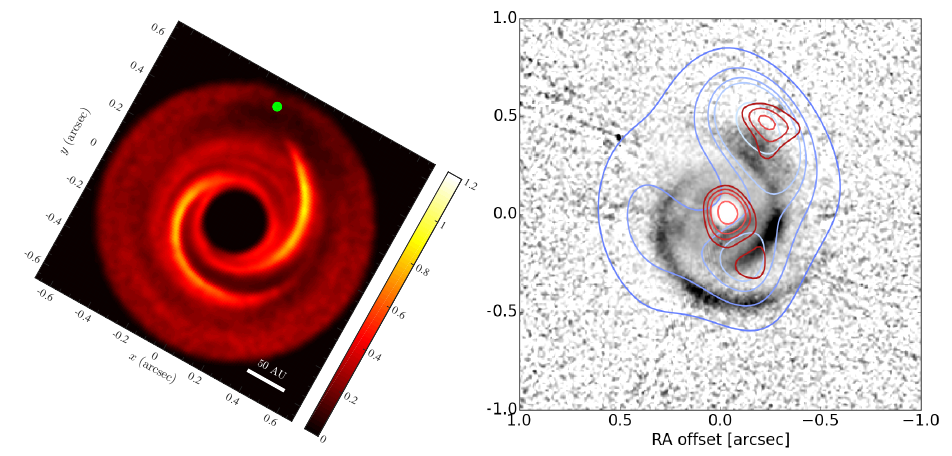}
\end{center}
\caption{ {\bf Left:} Spiral model involving a planetary mass
  companion, from \citet[][part of their Fig. 4, \copyright AAS,
    reproduced with permission]{Dong2015ApJ...809L...5D}. The
  companion is highlighted as a green dot - and the field is rotated
  so that the companion approximately matches the location of the VLA
  clump seen in MWC~758. {\bf Right} \citep[adapted
    from][]{Marino2015ApJ...813...76M}: VLA data, in red contours, and
  ALMA 345GHz, in blue contours, overlaid on the SPHERE Y-band
  polarised intensity image \citep{Benisty2015AA...578L...6B}, in grey
  scale.} \label{fig:mwc758spirals}
\end{figure*}

The compact VLA signal in MWC~758 may also be looked at from an
entirely different perspective. Another interpretation could follow
from the recent proposal by \citet{Dong2015ApJ...809L...5D} that the
observed spiral pattern in MWC~758 are launched by planet could be
launched by a planet exterior to the arms, at a radius of
$\sim$0.6~arcsec. \citet{Marino2015ApJ...813...76M} caution that the
observed signal is quite close to the location of such a body, as
illustrated in Fig.~\ref{fig:mwc758spirals}. Thus, the compact signal
could be circumplanetary dust that is heated by planet formation
feedback, \citep[as in the positive feedback leading to enhanced
  viscous heating proposed by ][]{Montesinos2015ApJ...806..253M}.

\subsection{Mild asymmetries and radial dust traps}

The above examples for extreme submm continuum asymmetries are seen in
few HAeBe stars (stellar mass 2~M$_\odot$), and are not typical of the
average TD.  Most TDs show milder azimuthal
asymmetries when resolved in the sub~mm continuum.  Examples of mild
contrast ratios, $\lesssim$4 or as could be reached by the gas, are
HD~135344B, SR~21 \citep[][]{Perez_L_2014ApJ...783L..13P,
  Pinilla2015A&A...584A..16P} and LkH$\alpha$~330
\citet[][]{Isella2013ApJ...775...30I}. Given their coarseness, these
observations could accommodate sharper asymmetries.

An interesting avenue to explain such mild asymmetries could be spiral
arm crowding. Indeed the asymmetry reported in HD~135344B
\citep[][]{Perez_L_2014ApJ...783L..13P} with a relatively coarse beam
in ALMA Cycle~0, has been resolved into a convergence of spiral arms
\citep[][their Fig.~1]{vdMarel_2016A&A...585A..58V}. These spirals
were hinted at in axisymmetric model residuals
\citep[][]{Perez_L_2014ApJ...783L..13P,
  Pinilla2015A&A...584A..16P}. Local shock heating from the spiral
waves \citep[][]{Rafikov2016arXiv160103009R} could perhaps further
pronounce asymmetries related to spirals, and lead to mild asymmetries
when observed in a coarse beam.

There are also examples of axially symmetric TDs, such as the TTauri
J160421, SZ~91, DoAr~44 \citep[][]{Zhang2014ApJ...791...42Z,
  vdMarel_2016A&A...585A..58V} and SZ~91
\citep[][]{Canovas2016arXiv160106801C}. While azimuthal dust
segregation or trapping does not seem to be required in such systems
with no or mild asymmetries, on the other hand the radial segregation
of dust grain sizes seems to be systematically present in TDs.  Radial
trapping is often invoked as a means to halt the catastrophic inward
migration of grains due to aerodynamic drag
\citep[][]{W1977MNRAS.180...57W}. The smaller and shallower cavities
in the gas phase compared to the dust were introduced in
Sec.~\ref{sec:gas} as a feature of dynamical clearing by giant planet
formation. This is the same phenomenon as radial trapping, since it is
the radial bump in gas pressure generated by planet formation that
traps the larger dust grains in the outer disk
\citep[e.g.][]{Pinilla2012AA...545A..81P, Pinilla2015A&A...584A..16P}.

\subsection{Near-term prospects: the gas background and vortex velocity fields}

The dust trap phenomenon is clearly at work in the two most extreme
asymmetries, and could be an important mechanism in the evolution of
the dust grain population. However, accurate knowledge of the gas
background and physical conditions is required to estimate the
efficiency of trapping by aerodynamic coupling. As yet there has been
no confirmation of the velocity fields expected in large-scale
vortices, so an important unknown on the origin of the lopsided gas
distributions remains untackled.  Another open question is, for
instance, the impact of enhanced cooling in the dust trap on the
maximum possible gas density contrast. 

Taking the dust trapping scenario one step further poses a question on
the potential of such dust traps to form gaseous giants. What would be
the impact on the disk of massive bodies forming inside a dust trap at
large stellocentric distances? Are there signs of such bodies?

Both the observational and theoretical situations are quickly
evolving.  ALMA should soon provide additional multi-frequency
continuum data with which to better quantify the degree of trapping
and the corresponding grain distributions. Perhaps resolved molecular
line maps of dust traps could be used to detect the expected vortex
velocity fields.  Without an observational confirmation, however, the
question on the origin of the gaseous lopsided structures is still
open.

\section{SPIRALS} \label{sec:spirals}

\begin{figure*}
\begin{center}
\includegraphics[width=\textwidth,height=!]{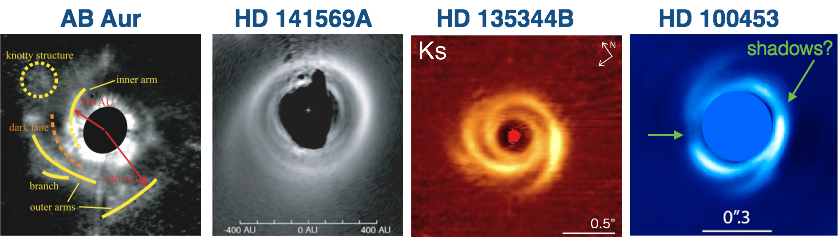}
\end{center}
\caption{ Gallery of optical/IR spirals.  From left to right, we show
  scattered-light images of AB~Aur \citep[][part of their Fig.~3,
    \copyright AAS, reproduced with
    permission]{Fukagawa2004ApJ...605L..53F} , HD~141569A
  \citep[][part of their Fig.~8, \copyright AAS, reproduced with
    permission]{Clampin2003AJ....126..385C}, HD~135344B \citep[][part
    of their Fig.~1, \copyright ESO, reproduced with
    permission]{Garufi2013AA...560A.105G}, and HD~100453 \citep[][part
    of their Fig. 2, \copyright AAS, reproduced with
    permission]{Wagner2015ApJ...813L...2W}.  All figures have been
  reproduced by permission of the AAS or A\&A. In HD~100453 we have
  highlighted two intensity dips where the two-armed spiral pattern
  seem to stem from. These dips are very reminiscent of the HD~142527
  shadows, which also seem to be at the root of
  spirals. } \label{fig:spirals}
\end{figure*}

Observations in optical/IR scattered light of spirals in
protoplanetary disks have been discussed on several occasions, both
in discovery articles (see below) and in modelling efforts
\citep[][]{Juhasz2015MNRAS.451.1147J, Pohl2015MNRAS.453.1768P,
  Dong2015ApJ...812L..32D}.  The single-star systems where spirals
have been reported seem to be TDs, i.e. they host either
central cavities or gaps. It is possible that these systems are in
fact binaries with a low mass ratio: their spiral patterns bear
similarities with the arc-like features seen in some binary disks,
such as in AS~205 \citep[][]{Salyk2014ApJ...792...68S} and SR~24
\citep[][]{Mayama2010Sci...327..306M}. The following is a list of
spiral patterns in disks around primary stars that dominate by mass
(relative to possible low mass bodies embedded in the disks), some of
which are shown in Fig.~\ref{fig:spirals}:

\begin{itemize}
\item HD~100546, with a 13~AU gap
  \citep[e.g.][]{Tatulli2011AA...531A...1T, Avenhaus2014ApJ...790...56A},
  and tightly wound and complex spirals seen in the outer disk
  \citep[][]{Grady2001AJ....122.3396G, Boccaletti2013AA...560A..20B};
\item HD~142527, with a 140~AU gap and both a grand design spiral
  pattern in the outer disk as well as more intricate spiral features
  carving the outer edge of the cavity \citep[][]{Fukagawa2006,
    Rameau2012AA...546A..24R, Casassus2012ApJ...754L..31C, Canovas2013A&A...556A.123C,
    Avenhaus2014ApJ...781...87A};
\item AB~Aur, with a gap between 40 and 140~AU \citep[][]{
  Hashimoto2011ApJ...729L..17H}, also exhibits large scale spirals
  that modulate the outer disk, and seen in scattered light
  \citep[][]{Grady1999ApJ...523L.151G, Fukagawa2004ApJ...605L..53F}; 
\item HD~141569A, in the transition to debris disk
  \citep[e.g.][]{Thi2014AA...561A..50T}, with several tightly-wound
  concentric rings or spirals extending out to 400~AU, and a 175~AU
  cavity \citep[][]{Biller2015MNRAS.450.4446B}, also hosts large-scale
  open spirals that may result from interaction with the other members
  of this hierarchical multiple system
  \citep[][]{Clampin2003AJ....126..385C};
\item MWC~758, with a central 100~AU cavity detected in the sub~mm
  \citep[][]{Isella2010ApJ...725.1735I} but not in the near-IR
  \citep[][]{Grady2013ApJ...762...48G}, shows a large-scale grand
  design two-armed spiral \citep[][]{Grady2013ApJ...762...48G,
    Benisty2015AA...578L...6B} recently interpreted as interior
  spirals triggered by a companion at fairly wide separation
  \citep[0.5~arcsec, see
    Fig.~\ref{fig:mwc758spirals},][]{Dong2015ApJ...809L...5D};
\item HD~135344B has a $\sim$30~AU dust cavity
  \citep[][]{Carmona2014AA...567A..51C}, and a 2-armed grand-design
  spiral best seen with polarised differential imaging
  \citep[][]{Muto2012ApJ...748L..22M, Garufi2013AA...560A.105G,
    Wahhaj2015AA...581A..24W};
\item HD~100453 is a recent addition to the list of known spirals in
  TDs
  \citep[][]{Wagner2015ApJ...813L...2W,Dong2016ApJ...816L..12D}.  This
  HAe disk, with ring enclosing a cavity of $\sim$20~AU, displays a
  two-armed spiral pattern stemming from the ring, fairly open and
  extending out to $\sim$42~AU.
\end{itemize}

\begin{figure*}
\begin{center}
\includegraphics[width=0.7\textwidth,height=!]{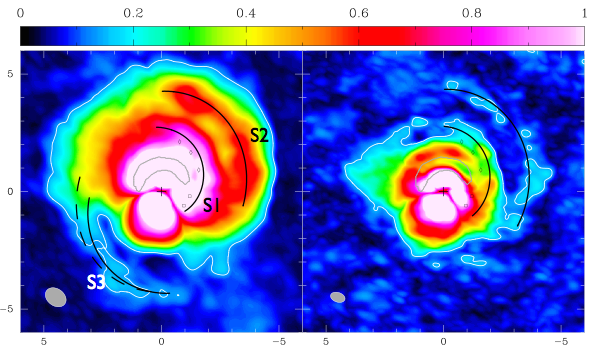}
\end{center}
\caption{Radio spirals from HD~142527 \cite[Figure adapted
    from][]{Christiaens2014ApJ...785L..12C}.  The CO(2-1) peak signal
  is shown on the left, while CO(3-2) is shown on the right (the gray
  ellipse correspond to the clean beams). $x-$ and $y-$axis indicate
  angular offset from the star, along RA and Dec., and in
  arcsec. Since the distance to HD~142527 is about 140~pc, the spirals
  are seen to extend out to 700~pc in radius. The lozenge symbols
  indicate the most conspicuous IR spiral, from
  \citet{Fukagawa2006}. The inverted-V decrement in peak CO intensity,
  seen in both transitions, is due to interstellar absorption at
  velocities that correspond to these locations in the Keplerian outer
  disk. } \label{fig:HD142spirals}
\end{figure*}

There are, however, only two examples of molecular-line counterparts
to the near-IR spirals in protoplanetary disks. The AB~Aur IR spirals
were also detected in the CO isotopologues
\citep[][]{Corder2005ApJ...622L.133C,
  Lin2006ApJ...645.1297L}. Intriguingly the spirals in AB~Aur appear
to counter-rotate with the disk, which led
\citet{Tang2012AA...547A..84T} to propose that they stem from an
infalling envelope (but the apparent counter-rotation could also
result from a Doppler flip due to a warp, see
Sec.~\ref{sec:gasrichwarps}). As summarised in
Fig.~\ref{fig:HD142spirals}, \citet{Christiaens2014ApJ...785L..12C}
reported very large scale $^{12}$CO spirals in HD~142527, one of which
appears to have a clear counterpart in the most conspicuous IR spiral
found by \citet{Fukagawa2006}.

The radio detection of spirals in molecular-lines is very important to
estimate physical conditions, and search for kinematic trends, such as
deviations from Keplerian rotation (as in AB~Aur), that may help
understand the origin of the spirals. Velocity dispersion maps may
perhaps also be used to detect the compression wave. For now, the
molecular-line spirals in HD~142527 allow estimates of temperature and
column densities. \citet{Christiaens2014ApJ...785L..12C} find that the
spirals are surprisingly cold, with 25~K near the continuum ring, and
down to 10-15~K further out. The Toomre~Q parameter, $Q = c_s \Omega
/ (\pi G \Sigma) $, for each of the two spirals $S_1$ and $S_2$, are $
100 < Q_{S1} < 50000$, and $50 < Q_{S2} < 35000$, where the lower
bound stems from the continuum non-detection (assuming a gas-to-dust
ratio of 100), and the upper bound is set by the assumption of
optically thin $^{12}$CO. Thus, the spirals seen in the outer disk of
HD~142527 appear to be gravitationally stable.

\subsection{Near-term prospects: radio counterparts}

The near future should see a multiplication of radio counterparts to
the IR spirals \citep[e.g. the continuum counterparts in
  HD~135344B][]{vdMarel_2016A&A...585A..58V}. Radio observations,
especially in molecular lines, are required to estimate physical
conditions such as density and temperature, and so find clues as to
the origin of spirals.  In addition radio observations should help in
disentangling radiative transfer effects, such as proposed by
\citet{Quillen2006ApJ...640.1078Q}.

\begin{figure}
\begin{center}
\includegraphics[width=\columnwidth,height=!]{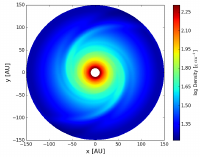}
\end{center}
\caption{ Spirals launched from the temperature forcing of the outer
  disk by shadows projected from a tilted inner disk - as in HD~142527
  (and perhaps also in HD~100453). The image shows a snapshot of the
  density field at 250~orbits, under the effect of shadows aligned in
  the East-West direction \citep[from ][]{Montesinos2016arXiv160107912M}.  } \label{fig:spiralshadows}
\end{figure}

There are several on-going efforts on the theoretical side, with the
coupling of 3D hydrodynamics and 3D radiative transfer applied to
planet-disk interaction \citep[e.g.][]{Ober2015A&A...579A.105O,
  Perez2015ApJ...811L...5P, Dong2015ApJ...809L...5D}. However, in
addition to the gravitational interaction with massive bodies, either
planetary-mass or stellar, as in multiple stellar systems, perhaps
other physical mechanisms may also launch spirals in circumstellar
disks. One possibility is gravitational instability of the disk, whose
observability in scattered light has recently been predicted from 3D
radiative transfer and hydrodynamics
\citep[][]{Dong2015ApJ...812L..32D}. Another example, motivated by the
obvious link of the HD~142527 spirals with the shadows cast on the
outer disk by the inner warp (which may also be observed in
HD~100453), is the proposal by \citet{Montesinos2016arXiv160107912M}
that spirals could result from the forcing of the temperature field in
the outer disk, as it is modulated by variable illumination due to
non-axially symmetric shadowing from the inner regions. This idea is
illustrated in Fig.~\ref{fig:spiralshadows}.

\section{THE ROLE OF EMBEDDED PROTOPLANETS} \label{sec:planets}

Key to understanding the astrophysics of circumstellar disks, and
eventually giant planet formation, is definitive information on the
proto-planetary systems that these disks potentially host.  Three
promising protoplanet detections have recently been reported from
their thermal signal at $\sim$3~$\mu$m, in the gapped disks HD~100546
\citep[][]{Quanz2013ApJ...766L...1Q, Quanz2015ApJ...807...64Q},
HD~169142 \citep[][]{Reggiani2014ApJ...792L..23R} and in LkCa~15
\citep[][]{Sallum2015Natur.527..342S}. In the case of LkCa~15 the
identification of an accreting  compact body is strengthened by the
detection of coincident and unresolved H$\alpha$ emission.

The interpretation of unresolved continuum signals at low contrasts is
difficult, as the protoplanet accretion luminosity leads to a positive
feedback in stellocentric accretion
\citep[][]{Montesinos2015ApJ...806..253M}, resulting in an extended
thermal signal well in excess of genuine photospheric emission, and
perhaps also a local enhancement in scale height. An example of such
extended near-IR emission, both in thermal emission (e.g. in $L$ band)
and in scattered-light (e.g. in $K$ band), could be the spatially
extended signal from HD~100546b \citep[][]{Garufi2016arXiv160104983G}.

In general the morphological details of the extended signal from such
candicate circumplanetary regions are very difficult to ascertain
because of observational biases, either in the high-contrast
techniques or in near-IR interferometry, which are being tackled by
technological breakthroughs (i.e. with extreme AO cameras such as GPI
and SPHERE, and the next-generation interferometry instruments such as
VLTI-MATISSE). Examples of such biases are the radiative transfer
effects at finite inclination that lead to asymmetries in otherwise
axially symmetric disks, and to similar closure phase signals as
binary models \citep[as in FL~Cha and
  T~Cha,][]{Cieza2013ApJ...762L..12C, Olofsson2013A&A...552A...4O}.

A different avenue to detect embedded protoplanets, and estimate their
mass, is to identify the corresponding kinematical signature of the
circumplanetary disk (CPD). \citet{Perez2015ApJ...811L...5P} study the
observability of such CPDs with ALMA using molecular lines, by
coupling 3D hydrodynamical simulations and radiative transfer, and
after filtering for the instrumental response\footnote{including real
  phase noise taken from the HL~Tau dataset presented by 
  \citet{2015ApJ...808L...3A}}. They predict that the CPD from an
accreting gaseous giant should stand out in velocity dispersion
maps. In addition the channel maps with disk emission at the position
of the protoplanet should show a kink (i.e. a tight local bend), whose
amplitude depends on the planet mass. Starting from ALMA Cycle~3, such
long baseline observations with ALMA are routinely possible, so that
prospects are good for the detection of the kinematical signatures of
CPDs.

\section{CONCLUSION} \label{sec:conclusion}

While accurate knowledge on the existence of embedded protoplanetary
systems is still far from systematically available, resolved
observations of TDs reveal surprising structures. The observed
phenomena do not readily fit in the picture expected for TDs as
sculptures of planet formation. Instead, the basic idea of flat
protoplanetary disk system is being challenged by the observations of
warps, whose frequency remains to be assessed.  Residual gas inside
cavities is clearly observed, as expected for dynamical clearing, but
the concomitant protoplanetary streamers are still elusive. The
observed intra-cavity stellocentric accretion, in the single case
where it has been detected, occurs at much higher velocities than
expected for planetary accretion streams.

The outer regions of TDs often seem to be non-axially symmetric -
quite dramatically so in a couple of instances. This observation could
bring confirmation for the scenario of `dust trapping' as a way to
circumvent the catastrophic infall of the larger dust grains due to
aerodynamical drag.  While dust size segregation has clearly been
demonstrated to exist, the origin of the required gaseous asymmetry in
a large anticyclonic vortex remains to be confirmed
observationally. There does not seem to be a clear observational
connection between the outer disk asymmetry and on-going planet
formation, as illustred by the case of LkCa~15, with a detection of
protoplanetary accretion \citep[][]{Sallum2015Natur.527..342S}, and a
seemingly perfectly symmetric outer disk
\citep[][]{vandermarel_2015AA...579A.106V}.

The observations of TDs have also revealed spectacular
spiral arms, the origin of which still remains the subject of intense
debate. They could be launched by planets, or result from
gravitational instability, or be induced from variable illumination
\citep[][]{Montesinos2016arXiv160107912M}.  Once launched, the
frequent observation of spirals in systems with cavities and
non-axially symmetric outer disks suggests that spirals may play a
crucial role in disk evolution, as argued by
\citet[][]{Rafikov2016arXiv160103009R}.  New data will soon provide
estimates of physical conditions with which to test these ideas.

Finally, it appears that physical processes other than planet
formation are governing the TD phase of Class~II YSOs, since most
large cavities seem empty of massive bodies. Perhaps well identifiable
planet formation structures have to be searched for at earlier
evolutionary stages, i.e.  in Class~I disks such as HL~Tau
\citep[][]{2015ApJ...808L...3A, Testi2015ApJ...812L..38T,
  Dipierro2015MNRAS.453L..73D, Gonzalez2015MNRAS.454L..36G,
  Pinte2016ApJ...816...25P}.


\begin{acknowledgements}
  We thank the following disk workers for very useful input: Henning
  Avenhaus, Amelia Bayo, Mark Booth, Valentin Christiaens, Lucas
  Cieza, Sebasti\'an Marino, Mat\'{\i}as Montesinos, Johan Olofsson,
  Sebasti\'an P\'erez, Dave Principe, Nienke van der Marel, Gerrit van
  der Plas and Matthias Schreiber. Financial support was provided by
  Millennium Nucleus RC130007 (Chilean Ministry of Economy), and
  additionally by FONDECYT grant 1130949.
\end{acknowledgements}

\bibliographystyle{pasa-mnras}

\input{diskobs.bbl}


\end{document}